\documentclass[a4paper,fleqn]{cas-dc}

\usepackage[numbers]{natbib}

\usepackage{tabularx}
\usepackage{url}
\usepackage{listings}
\usepackage{xcolor}

\lstdefinelanguage{Turtle}{
  sensitive=true,
  alsoletter={@},
  morekeywords={@prefix,@base,PREFIX,BASE,a,true,false},
  morecomment=[l]{\#},
  morestring=[b]",
  morestring=[s]{<}{>},
}

\def\tsc#1{\csdef{#1}{\textsc{\lowercase{#1}}\xspace}}
\tsc{WGM}
\tsc{QE}
\begin{document}
\let\WriteBookmarks\relax
\def\floatpagepagefraction{1}
\def\textpagefraction{.001}

\shorttitle{AI Risk Ontologies for the EU AI Act Compliance}    

\shortauthors{D. Golpayegani et al.}  

\title [mode = title]{AI Risk Ontologies for the EU AI Act Compliance: AIRO, VAIR, and their Integration with DPV}

\author[1]{Delaram Golpayegani} [orcid=0000-0002-1208-186X]
\cormark[1]
\ead{sgolpays@tcd.ie}

\credit{Conceptualization, Methodology, Software, Validation, Investigation, Visualization, Writing - Original Draft, Writing - Review \& Editing}

\author[1]{Harshvardhan J. Pandit} [orcid=0000-0002-5068-3714]
\ead{me@harshp.com}
\credit{Conceptualization, Writing - Review \& Editing, Supervision}

\author[1]{Declan O'Sullivan} [orcid=0000-0003-1090-3548]
\ead{declan.osullivan@tcd.ie}
\credit{Writing - Review \& Editing, Supervision}

\author[1]{Dave Lewis} [orcid=0000-0002-3503-4644]
\ead{delewis@tcd.ie}
\credit{Conceptualization, Writing - Review \& Editing, Supervision, Funding acquisition}

\affiliation[1]{organization={ADAPT Centre,  School of Computer Science and Statistics, Trinity College Dublin},
  %          addressline={}, 
  %          city={},
%          citysep={}, % Uncomment if no comma needed between city and postcode
            country={Ireland}}

\cortext[cor1]{Corresponding author}



\begin{abstract}
As the use of AI, and consequently its harms, becomes more prevalent, the global discourse around AI has become increasingly risk-aware. 
With new risks emerging more frequently than ever, a key challenge is how to represent, document, and catalogue AI risks in an effective, interoperable, and queryable manner. Additionally, the European Union (EU) AI Act's requirements for``high-risk'' AI systems to manage and document the risks creates the need for open and interoperable tools, methods, and standards that facilitate AI risk management in an ever-changing environment. In this paper, we present ontologies and taxonomies developed to assist with risk management and documentation, in line with the requirements of the EU the AI Act. We developed the AI Risk Ontology (AIRO) to specify AI use cases and their associated risks, complemented by the Vocabulary of AI Risks (VAIR), which populates AIRO with taxonomies of concepts. Key international standards on AI and risk, namely ISO/IEC 23894 on AI risk management and ISO/IEC 22989 on AI terminology informed the development of the ontologies. To promote reuse and enable multi-regulatory compliance, AIRO and VAIR have been integrated in the Data Privacy Vocabulary (DPV) through a consensus-building process, expanding DPV's scope from data to AI.

\end{abstract}



%\item 

% Keywords
\begin{keywords}
AI Risk \sep EU AI Act \sep Ontology \sep AI governance
\sep Legal compliance \sep ISO standard

\end{keywords}

\maketitle

\section{Introduction}\label{s:intro}
With the growing body of evidence on harms caused by Artificial Intelligence (AI), the assertion that the use of AI involves risk is no longer contested, though there are different school of thoughts on what constitutes AI risk. The global AI discourse is becoming increasingly risk-aware, but still regulating AI is an open question for some jurisdictions prioritising innovation-first views. The European Union (EU) is among those that have taken a precautionary approach towards AI and has become the first jurisdiction that adopted a regulatory framework for AI, called the AI Act~\cite{eu-ai-act}. By taking a risk-based approach, the AI Act aims to protect individuals from a broad range of potential AI harms, ranging from risks to health and safety to infringements of fundamental human rights. 

Compliance with requirements of the AI Act, including risk management (Article 9) and technical documentation (Article 11), require large volumes of information to be maintained, queried, inferred, and shared. This is particularly challenging as this information should reflect the changes to the system, its components, and its context, all of which may evolve too quickly and frequently. Compounding this challenge, the AI Act itself is subject to change. The Act was enacted in 2024, after 3 years of negotiations between EU institutional triangle—the European Commission, the European Parliament, and the Council of the European Union. Although not all obligations under the AI Act are yet into force (as of July 2026), a set of technical amendments to the AI Act as a part of the \textit{Digital Omnibus Package} to simplify the EU digital acquis has been entered into force in July 2026~\cite{ominbus-aiact}. In addition, the AI Act includes mechanisms to update the regulation, such as amendments to the list of high-risk applications, through delegated acts.  While these amendments are intended to keep the AI Act aligned with the rapidly evolving AI capabilities, they introduce \textit{uncertainty} concerning the legal requirements and the scope of information that should be maintained for ongoing compliance. 

Given the importance of AI risks, both within the regulatory framework of the EU AI Act and in the broader context of safe and trustworthy AI, this paper leverages Semantic Web technologies to develop an ontology for describing AI risks, in an interoperable, machine-readable manner. Since AI risk is context-dependent, the ontology aims to provide a semantic model for specifying both \textbf{AI use cases} and their \textbf{risks}. Ensuring extensibility and reusability, the ontology is designed as a high-level, horizontal semantic model, meaning it is not tied to a specific AI technology, application domain, or sector. In addition to adaptability to changes, the ontological approach enables semantic consistency, which is particularly important as the implementation of the AI Act relies on \textit{harmonised standards} (Article 40). These standards are being developed by European Standardisation Organisations (CEN and CENELEC) and in terms of terminology and definitions are not necessarily fully consistent with the AI Act. Further, utilising Semantic Web technologies ensures machine-readability and in turn interoperability, supporting effective information sharing across the AI value chain and with authorities for conformity assessment purposes. %Semantic Web technologies have also been successfully adopted for implementation of RegTech (Regulatory Technology) solutions, in particular for compliance with the EU digital regulations (cf. ). %Also the European Commission's EuroVoc and SEMIC.

%reusable manner - logical consistency -

 In this paper, we develop  AI Risk Ontology (AIRO) as a FAIR (Finable, Accessible, Interoperable, Reusable)~\cite{wilkinson2016fair} ontology that enables modelling of AI use cases and their risks, in line with the requirement of the EU AI Act. We further develop the Vocabulary of AI Risks (VAIR), as a specialisation of AIRO to include hierarchies of concepts based on the EU AI Act and ISO/IEC standards related to AI risks, including ISO/IEC 22989 on AI concepts and terminology~\cite{iso22989}, ISO/IEC 23894 on AI risk management~\cite{iso23894} and ISO 31073 on risk management vocabulary~\cite{iso31073}. To promote creation and maintenance of standards-based and interoperable solutions, we proposed AIRO and VAIR to be integrated into the Data Privacy Vocabulary (DPV)~\cite{pandit2024dpv}, as the state-of-the-art vocabulary grounded in the EU digital regulations and a sustained W3C community resource, through a consensus-driven process within the DPV Community Group (DPVCG). By incorporating AIRO and VAIR, DPV's scope has been extended beyond data protection to include AI and EU AI Act. The overall methodology, which involved designing the ontologies while the AI Act was still under development and their integration with DPV represents an additional contribution of this paper, providing a manageable and modular approach in constructing consistent and integrable vocabularies for AI risks. 

The remainder of the paper is structured as follows: 
\autoref{s:stoa} reviews the state-of-the art ontologies and taxonomies for AI risks. \autoref{s:methodology} discusses the ontology engineering methodology, with a focus on dealing with changing legal requirements. Moreover, it provides the best practice to integrate an already-developed ontology into a standardised ontology through a consensus-building process. \autoref{s:airo} and \autoref{s:vair} describe AIRO and VAIR, respectively, with their evaluation provided in \autoref{ss:eval-quality}. \autoref{s:dpv_integration} discusses how AIRO and VAIR are integrated into DPV. Since initial version of AIRO and VAIR have been published in our earlier work~\cite{golpayegani2022airo, golpayegani2023high}, in \autoref{s:impact} we review their uptake both by AI researchers and industry practitioners. Finally, \autoref{sec:conclusion} concludes the paper.

\section{State of the Art} \label{s:stoa}

AI risk is getting more attention than it ever has, and consequently, there has been a surge in studies that identify, analyse, and classify such risks. According to a recent study by MIT FutureTech, at least 74 frameworks for AI risk exists~\cite{slattery2026ai}. Without aiming to be exhaustive, in this section, we provide an overview of emerging body of AI risk taxonomies and ontologies. In this overview, acknowledging the growing body of work on specific risk taxonomies, e.g. sector- or technology-specific ones, we emphasis on \textit{generic} AI risk ontologies, given the scope of the objectives that this paper pursues. Further, as will be shown, despite the vast body of work, most of AI risk taxonomies have been disseminated through research papers, without adopting Linked Open Data principles. However, such taxonomies are included due to the influence they already exerted.

\subsection{Taxonomies for Cataloguing AI Incidents}

With the rise of incidents caused by AI, there have been multiple initiatives focused on cataloguing these incidents within repositories. To serve a diverse set of stakeholders, such repositories usually provide incidents in a structured and annotated format. Therefore, many of existing generic taxonomies for describing AI systems and their risks have grown out of the need to annotate AI incidents. 
This subsection refers to the most prominent taxonomies derived from or associated with AI incident repositories.

The \textbf{OECD framework for classification of AI systems}~\cite{oecd2022classificationframework}, is a tool for assessing potential risks and benefits of AI use cases by considering five high-level dimensions: people \& planet, economic context, data \& input, AI model, and task \& output. The framework also incorporates taxonomies for its risk assessment criteria. 
The \textbf{AI, Algorithmic, and Automation Incidents and Controversies (AIAAIC)} repository\footnote{\url{https://www.aiaaic.org/aiaaic-repository}}, which is an open-access living dataset of more than 2200 AI incidents covered by the media, uses a taxonomy for annotating the incidents. The set of concepts for incident annotation includes categories of sectors, technologies, purposes, and impacts of AI on individuals, society, environment, and providers\footnote{\url{https://www.aiaaic.org/aiaaic-repository/classifications-and-definitions}, visited on 22/10/2026}. An extension of this taxonomy with a focus on AI harms has also been developed, wherein AI harms are classifies into 9 top-level categories and 69 sub-categories~\cite{abercrombie2024aiaaic} . 
\textbf{The Center for Security and Emerging Technology (CSET) AI harm framework} \cite{hoffmann2023csettaxonomy} is a generic customisable framework that has been developed based upon annotating AI incidents through a collaborative process. In its classification of harms, at the top-level it considers the following types of harms: \emph{tangible} and \emph{intangible}, and distinguishes between the \emph{actual} and \emph{reasonable potential} harm, referring to them as harm event and harm issue, respectively. 
The framework also enables linking AI harms to the associated AI system and harmed entities. The CSET AI Harm Framework was customised for annotating \textbf{AI incident database (AIID)}\footnote{\url{https://incidentdatabase.ai/}}~\cite{mcgregor2021aiid}
 with a fine-grained, yet not formal, schema provided in~\cite{hoffmann2023csetaiid}. 
 The \textbf{Goals, Methods, and Failures (GMF) taxonomy}~\cite{pittaras2023gmfaiid} has also been developed based on AIID and used for its annotation. 
The taxonomy considers three aspects of an AI incident: the AI system's goals, the technologies used within the system, and technical causes of the incident. 

The \textbf{taxonomy of AI Vulnerability Database (AVID)}\footnote{\url{https://avidml.org/database/}} \cite{avid} considers 3 dimension of risks, which are security, ethics, and performance, across the machine learning (ML) development lifecycle. The notable advantage of the AVID taxonomy is the JSON schema, providing better interoperability and extensibility compared to the other taxonomies discussed in this subsection. 

\subsection{AI Risk Taxonomies}

In regard to horizontal AI risk taxonomies, \textbf{Eticas AI Risk Taxonomy} \cite{clavell2026eticasairisktaxonomy} includes 10 high-level AI risks, 21 sub-groups representing risk assessment areas and further 70 sub-categories for test scopes. The Eticas taxonomy also includes a mapping to standards and frameworks, including the EU AI Act and ISO/IEC 42001, among others. \textbf{IBM's AI risk atlas taxonomy}~\cite{bagehorn2025airiskatlastaxonomy} organises AI risks based on where the risk originates, which are training data, inference, output, non-technical, and agentic. Inspired by our previous work, both taxonomies are encoded in machine-readable formats. 

The MIT risk taxonomy~\cite{slattery2026ai} collates and organises 1,725 risks, extracted from 74 existing taxonomies and frameworks, into two taxonomies: Causal Taxonomy, with categorisation using the sources of the risk, and Domain Taxonomy which focuses on risk consequences and impacts. While the taxonomy is provided in csv format (spreadsheets), it is not represented in a machine-readable format.

 To acknowledge the body of work on taxonomies of AI risks with a scope limited to the type of AI system, e.g. LLMs, or the domain of use, e.g. health and without aiming to be exhaustive, in the remainder of this subsection we refer to some promising specific risk taxonomies, which are: Weidinger et al.'s taxonomy of ethical and social risks of LLMs \cite{weidinger2022lmtaxonomy}, Tanaka et al.'s taxonomy for risks of generative AI systems \cite{tanaka2024genai}, Lee et al.'s taxonomy of AI privacy risks \cite{lee2024privacyrisks}, the Open Loop's taxonomy of potential harms associated with machine learning applications and automated decision-making systems \cite{andrade2021openloop}, NIST's taxonomy for adversarial machine learning \cite{nist2024adversialml} and categories of AI Bias \cite{nist2022bias}, Steimers and Schneider's work on taxonomy of risk sources that impact AI trustworthiness~\cite{steimers2022risksources}, and Roselli et al.'s work on classification of AI bias~\cite{roselli2019bias}. It should be noted that these taxonomies have been disseminated in a textual format as a part of research papers, without using any standardised formats.

\subsection{AI Risk Ontologies}
At the time of writing, there is an absence of ontologies that specifically provide a \textit{holistic} model of AI risks. However, there are state of the art ontologies that address trustworthy AI issues. It should be noted that these ontologies were reviewed as a part of the ontology development process (\autoref{ss:airo-reuse}) for ontology reuse.

The \textbf{EA-Ontology}\footnote{\url{https://protect.oeg.fi.upm.es/eaontology/eaontology_widoco/index-en.html}} \cite{vazquez2022ea} has been designed for ethical assessment of emerging technologies, including AI, based on existing approaches for ethical analysis. Within its module for modelling \emph{levels of ethical analysis}, the concept of \texttt{ateo:Risk} enables modelling risks identified during the ethical assessment and further linking them to ethical issues. 

The \textbf{Ethical AI principles ontology (AIPO)} \cite{harrison2021eaiontology} is a semantic model for trustworthy AI principles emerged from different guidelines and policies. The ontology models principles as \texttt{skos:Concept}, which enables relating and matching principles through leveraging SKOS relations. Within AIPO, trustworthy AI documents are modelled as \texttt{dcat:Resource}, allowing modelling the details of the document from which the principles are extracted.

The \textbf{Ai System use
Case Explanation oNTology (ASCENT)} \cite{adhikari2022ascent} is a framework for describing explainable AI measures. The framework includes 3 aspects: (i) technical information about the system that potentially impact explain AI solutions, (ii) information regarding the context of use that is related to explainability requirements, and (iii) the characteristics of the explanation measures. Within the use case module, a \texttt{Risk Level} (high risk or low risk) is assigned to a task for which the AI system is used, as an indicator of the significance of consequences. 
\textbf{ExplainableMLOntology} \cite{nakagawa2021xai} is an ontology aiming to enhance explainability of ML models through a process-oriented viewpoint. The ontology focuses on expressing processes used in the ML development and explanation and includes 3 modules for representing general ML, ML classification, and explanation. 
\textbf{FIDES ontology}\footnote{\url{https://w3id.org/fides}} \cite{fernandez2023fides} is a minimal semantic model of accountability-related
information about statistical ML models. FIDES supports expressing ML models and their data across different lifecycle phases. 

The \textbf{Doc-BiasO Ontology} \cite{russo2024biaso} has been proposed for expressing and documenting biases across ML/AI  workflows. Doc-BiasO comprises a high-level ontology describing the core concepts and relations, with a controlled vocabulary extending the ontology. The \textbf{Artificial Intelligence Ontology (AIO)}\footnote{\url{https://bioportal.bioontology.org/ontologies/AIO}} \cite{joachimiak2024aio} has been developed using a hybrid approach which consists of manual ontology development and automated processes empowered by the capabilities of LLMs. AIO consists of 444 concepts related to AI, grouped into 6 modules: networks, layers, functions, LLMs, preprocessing, and bias. Despite being called an ontology, it lacks object properties to model the relations between concepts.

 The \textbf{ontology for standardising trustworthy AI} \cite{Lewis2021tai} is an ontology for representing and mapping trustworthy AI concepts from different emerging AI standards. The ontology provides a way to express activities related to manifestation of AI trustworthiness. Enhancing this ontology, the \textbf{Trustworthy AI Requirements Ontology (TAIR)}~\cite{hernandez2024tair} has been developed for mapping AI regulations and standards. \texttt{Requirement} is the central element in TAIR, which can be linked to the entities, activities, and agents involved in implementation of the requirement. It further supports modelling the relationships between requirements extracted from different AI documents, e.g. for demonstrating how the EU AI Act's requirements can be satisfied by standards. 

Another work on modelling requirements is  Sabaliauskaite et al.'s ontology for dual-use regulatory compliance~\cite{sabaliauskaite2026automated}, which  models regulatory compliance requirements for AI systems that can be used for both civil and military purposes. The ontology models requirements from both civilian and military deployment contexts, drawing on EU digital regulations, such as the AI Act and GDPR, for the former and NATO frameworks, e.g. NIAP, NSP, for the latter. 

%%Comparison

\section{Ontology Development Methodology} \label{s:methodology}
\subsection{Scope}\label{ss:scope}
The EU AI Act served as the primary knowledge resource for the development of AIRO and VAIR. Since the AI Act only specifies the \textit{essential requirements} and relies on the highly-anticipated \textit{harmonised standards} to assist with interpretation of the Act and provide details for its implementation. At the time of writing, the full list of harmonised standards are not yet published although the deadline for the publication of such standards has passed. To assist with  the development of the ontologies, we used  ISO/IEC standards relevant to AI risks that are strong candidate for harmonisation, as additional knowledge sources. 

Over the course of undertaking this work, the AI Act was undergoing the EU ordinary legislative process \cite{eutreaty}. As a result of this process, multiple versions of the Act were published by EU authorities from April 2021 until July 2024. Consequently, this work has evolved over 7 distinct iterations of the AI Act, as listed in the following in chronological order: 
\begin{enumerate}
    \item The European Commission's proposal \cite{aia-commission}, published in April 2021,
    \item The French Presidency of the Council of the EU's Consolidated text \cite{aia-fr}, published in June 2022,
    \item The Czech Presidency of the Council of the EU's General Approach \cite{aia-cz}, published in November 2022, 
    \item The European Parliament's mandate \cite{aia-ep}, published in June 2023,
    \item The provisional agreement \cite{aia-provisional} resulting from inter-institutional negotiations during the trilogue phase, published in February 2024, 
    \item The Corrigendum \cite{aia-corrigendum}, published in April 2024,
    \item The final text \cite{eu-ai-act}, published in Official Journal of the EU in July 2024. 
\end{enumerate}

The AI Act introduces requirements for both AI systems and general-purpose AI models. However, this work focuses on AI systems as there is a divergence between risk management process for AI systems and general-purpose AI models under the AI Act. As the AI Act's regulatory regime is structured around risk, the obligations related to the risk-based classification and risk management are central in compliance with the Act's requirements for high-risk AI systems. Reflecting this emphasis on risks in the development of ontologies, we focused on risk management, documentation, and information sharing requirements referred to in the articles and annexes listed in \autoref{tab:<ch1-articles>}.

\begin{table}[!h]
    \centering
    \caption{List of articles and annexes from the AI Act \cite{eu-ai-act} used as key sources in development of the ontologies}
    \small
    \begin{tabularx}{0.5\textwidth}{|p{2cm}|X|}
    \hline
       \textbf{Obligations related to}  & \textbf{Sources from the AI Act}    \\
   \hline
   
    Risk management &  
    Article 5 - Prohibited AI practices \newline
    Article 6 - Classification rules for high-risk AI systems \newline
    Annex I - List of Union harmonisation legislation \newline
    Annex III - High-risk AI systems referred to in Article 6(2) \newline
    Article 9 - Risk management system\\

   \hline
    AI and risk documentation  & Article 9 - Risk management system \newline 
    Article 11 - Technical documentation
    \newline
    Annex IV - Technical documentation referred to in Article 11(1)
    \\
   \hline
    AI and risk information sharing & 
    Article 13 - Transparency and provision of information to deployers \newline
    Article 49 - Registration \newline
    Article 71 - EU database for high-risk AI systems listed in Annex III  \newline
    Annex VIII - Information to be submitted upon the registration of high-risk AI systems in accordance with Article 49
    \\
    \hline
    \end{tabularx}
    \label{tab:<ch1-articles>}
\end{table}

Regarding standards, CEN-CENELEC Joint Technical Committee 21 (JTC 21) is working with the European Commission to deliver the requested harmonised standards. This means that harmonised standards in relation to the AI Act are not published yet. In this absence, the ISO/IEC standards listed in \autoref{tab:<ch1-standards>}, which are strong contenders for harmonisation, are used in development of the ontologies. %It, however, should be noted that these standards have been found insufficient in meeting the requirements of the AI Act~\cite{soler2023analysis}.

\begin{table}[!h]
    \centering
    \caption{List of standards used as key sources to assist with interpretation of the AI Act}
    \small
    \begin{tabularx}{0.5\textwidth}{|p{2cm}|X|}
    \hline
       \textbf{Obligations related to}  & \textbf{Standards}    \\
   \hline 
    Risk management & ISO/IEC 23894:2023 - Artificial intelligence — Guidance on risk management \cite{iso23894} \newline
    ISO 31000:2018 - Risk management — Guidelines \cite{iso31000} \newline
    ISO 31073:2022 - Risk management — Vocabulary \cite{iso31073}   
   \\
   \hline
    AI and risk documentation and sharing  &
    ISO/IEC 22989:2022 - Information technology — Artificial intelligence — Artificial intelligence concepts and terminology \cite{iso22989} \newline
    ISO/IEC 42001:2023 - Information technology — Artificial intelligence — Management system \cite{iso42001}
 
    \\
    \hline
    \end{tabularx}

    \label{tab:<ch1-standards>}
\end{table}

\subsection{Methodology} \label{s:<overall-methodology>}

The development of both AIRO and VAIR followed the Linked Open Terms (LOT) methodology~\cite{poveda2022lot}—a lightweight methodology for developing ontologies and vocabularies in an iterative manner and aligned with industrial development. In creation of the ontologies, we also used Noy and McGuinness's ``Ontology Development 101'' guideline~\cite{noy2001ontology} to follow best practices.

\begin{figure*}[ht]
    \centering
    \fbox{\includegraphics[width=0.60\textwidth]{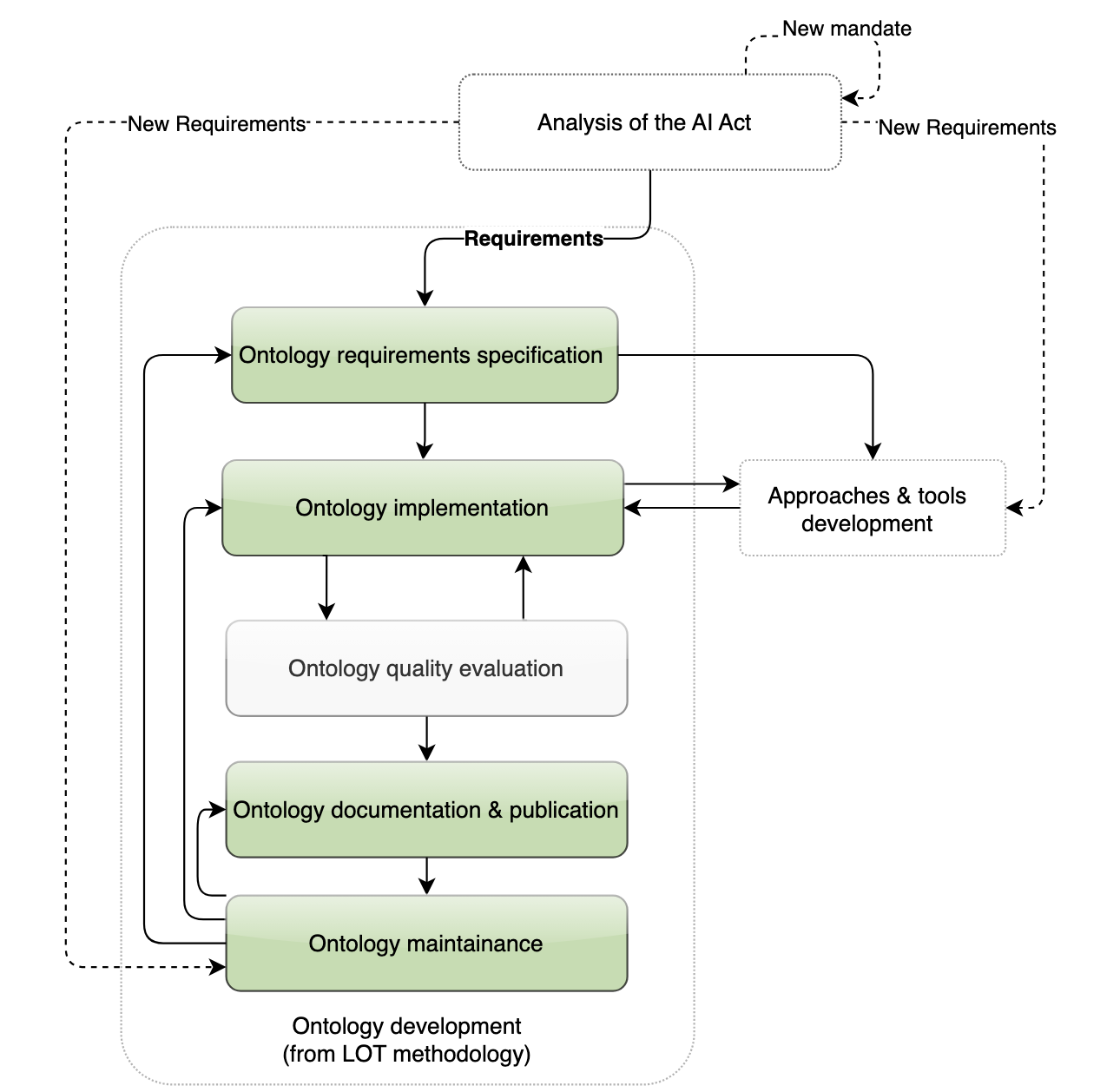}}
    \caption{Ontology development methodology}
    \label{fig:<ch4-ontologydev-methodology>}
\end{figure*}

%To deal with multiple mandates of the AI Act, the methodology for this research enabled frequent amendments to the ontologies. 

 In development of AIRO and VAIR we followed the four core activities of LOT, while considering additional iterative loops to deal with changing requirements arising from multiple mandates of the AI Act. In addition, due to the importance of ontology evaluation, we consider it as a separate activity from ontology implementation. The steps followed for development of the ontologies are illustrated in \autoref{fig:<ch4-ontologydev-methodology>} and is described in the following:
\begin{enumerate}
    \item \emph{Ontology requirements specification}: Based on our analysis of the AI Act, provided in our previous work, the functional requirements are specified in 4 sets of competency questions addressing (a) identification of 
    high-risk AI systems~\cite{golpayegani2023high}, (b) describing intended purpose of AI systems~\cite{golpayegani2024aiup}, (c) documentation of AI and risk information~\cite{golpayegani2024AIcards}, and (d) registration of high-risk AI systems into the EU database~\cite{aicat}. These are defined based on the scope of our research and towards developing Semantic Web-based approaches and tools for compliance with the EU AI Act's risk management and documentation obligations. 
    
    \item \emph{Ontology implementation}: In this phase, the concepts, their definitions, and relations are primarily derived from the AI Act. As discussed in \autoref{ss:scope}, where the details required for satisfying the competency questions are not provided by the AI Act, the information modelling utilises ISO/IEC standards. To ensure transparency and traceability, the sources from which the definitions are derived are encoded using \texttt{dcterms:source}.
    The ontologies are encoded in OWL 2 Web Ontology Language~\cite{owl2} using Protégé ontology editor~\cite{musen2015protege}. Additionally, for representing concepts in VAIR, the W3C SKOS data model~\cite{skos} is used.  
    In the encoding process, existing vocabularies for describing metadata and existing relevant ontological resources are reused (see \autoref{ss:airo-reuse}). %It is worth mentioning that the ontologies reviewed in \autoref{s:stoa} were identified as improper for reuse, mainly due to the fact that they are not aligned with the AI Act.

     \item \emph{Ontology evaluation}: AIRO and VAIR are evaluated against the competency questions to ensure their sufficiency in addressing those questions. Their quality is ensured by following the W3C Data on the Web Best Practices~\cite{w3c_dataonweb}, and FAIR best practices for vocabularies and ontologies \cite{garijo2020fair}. They are tested iteratively using OOPS! (OntOlogy Pitfall Scanner!)~\cite{poveda2014oops} and FOOPS! (ontology pitfall scanner for FAIR principles)~\cite{garijo2021foops}, as will be discussed in \autoref{ss:eval-quality}. AIRO and VAIR are also evaluated through their application in modelling AI use cases and their associated risks (see \autoref{ss:poc}). In addition, their applicability in different compliance tasks is demonstrated in our further work on (a) determining Annex III high-risk AI systems~\cite{golpayegani2023high}, (b) expressing AI intended purposes within a use policy~\cite{golpayegani2024aiup}, (c) documentation of AI and risk information~\cite{golpayegani2024AIcards}, and (d) cataloguing AI systems for registration into the EU database~\cite{aicat}. %using  use cases%one synthetic use cases (proctify) and two AI incidents (Uber's Real Time ID and VioGén Domestic Violence System). 
     
    \item \emph{Ontology documentation and publication}: The HTML documentation is created in a semi-automated process using the ReSpec template\footnote{\url{https://www.w3.org/respec/}}— a common template for W3C specifications. WIDOCO \cite{garijo2017widoco}, a well-known tool for automated generation of ontology documentation, is not used due to its limitations in supporting grouping of taxonomies of concepts \cite{pandit2024dpv}.
    Both AIRO and VAIR use permanent URLs (PURLs) using the W3ID service\footnote{\url{https://w3id.org/}} and are made available online at \url{https://w3id.org/airo} and \url{https://w3id.org/vair} under the CC-BY-4.0 license\footnote{\url{https://creativecommons.org/licenses/by/4.0/deed.en}}, allowing free reuse, distribution, and modification conditioned upon appropriate attribution.
    Both ontologies are indexed in the Linked Open Vocabularies (LOV) registry\footnote{\url{https://lov.linkeddata.es/dataset/lov/}} and published in Zenodo for better findability and accessibility. 
    
     \item\emph{Ontology maintenance}: As mentioned earlier, during the time the development ontologies have been carried out, the AI Act was undergoing the legislative process and therefore the requirements and concepts derived from the Act needed revisions after publication of each mandate. Additionally, multiple drafts of the relevant ISO/IEC standards were published over the course of development of the ontologies, which to some extent influenced the ontology design. AIRO and VAIR were iteratively updated following the LOT methodology, with additional iteration cycles. While all the changes were recorded through GitHub commit history, multiple versions were not released, as the aim of this work was to provide a first stable released aligned with the final version of the AI Act, rather than modelling its legislative evolution.
     
\end{enumerate}

\section{AI Risk Ontology (AIRO)} \label{s:airo}

AIRO provides a formal representation of AI systems and their risks as per the EU AI Act. As emphasised before, the AI Act only defines essential requirements and relies on future harmonised standards and common specifications for supporting its interpretation and implementation. With no such standards available at the time of writing, AIRO uses existing international standards as sources to extract concepts and their relations. 

\subsection{Iterative Development of AIRO}

The first iteration of the development, in 2021, employed the following standards from ISO's well-known 31000 series on risk management: ISO 31000:2018 on risk management guidelines~\cite{iso31000} and ISO Guide 73:2009 on risk management vocabulary~\cite{isoguide73}, which was later replaced by ISO 31073:2022 Risk management — Vocabulary~\cite{iso31073}. This replacement required the sources (\texttt{dct:source}) to be updated, however it did not impact the design of AIRO. With official publication of ISO/IEC 22989~\cite{iso22989} in 2022, the AI concepts and definitions were revised. While the terms used within the AI Act and ISO/IEC 22989 are overlapping, in some cases definitions and taxonomical relations are not exactly the same. In the process of development, the AI Act served as the primary resource and therefore definitions from standards were used when there was a lack of definition for a specific concept in the AI Act. Another round of revision was conducted when ISO/IEC 23894~\cite{iso23894} on AI risk management was published in 2023. Given that this standard relies on the definitions provided in ISO 31000:2018, ISO 31073:2022, and ISO/IEC 22989:2022, AIRO's concepts and definitions did not require any changes to be aligned with ISO/IEC 23894. In addition to the aforementioned standards, the definitions provided in ISO/IEC TR 24028:2020 — Overview of trustworthiness in artificial intelligence~\cite{iso24028} were used for some of the AI-related concepts. Further, the definition for \texttt{Standard} was extracted from Regulation (EU) No 1025/2012 on European standardisation~\cite{eu-standardisation-reg}, as the AI Act relies on the definition of harmonised standards in this regulation. \autoref{tab:<ch4-airo-ontology-sources>} shows the non-ontological sources from which definitions for AIRO concepts were extracted. To ensure traceability, from the Dublin Core Metadata Terms~\cite{dcterms}, \texttt{dct:source} is used to link the concepts to the source wherein they are defined.

\begin{table*}[!h]
    \centering
    \small
    \caption{Non-ontological sources of AIRO}
    \begin{tabularx}{\textwidth}{|p{6cm}|p{1.3cm}|X|}
        \hline
       \textbf{Source}  & \textbf{No. of concepts} & \textbf{AIRO Concepts} \\
       \hline
      
        The AI Act, Article 3 & 5 & \texttt{AI System}, \texttt{AI Operator}, \texttt{AI Deployer}, \texttt{AI Provider}, \texttt{General Purpose AI Model}  \\
        \hline
        ISO 31073:2022~\cite{iso31073} & 8 &  \texttt{Risk Source}, \texttt{Hazard}, \texttt{Threat}, \texttt{Vulnerability},
        \texttt{Consequence}, \texttt{Risk Control}, \texttt{Frequency}, \texttt{Likelihood}  \\
        \hline
        
        ISO/IEC 22989:2022~\cite{iso22989} & 3 & \texttt{AI Component}, \texttt{Stakeholder}, \texttt{AI Developer}  \\
       \hline
        ISO/IEC TR 24028:2020~\cite{iso24028} & 2 &  \texttt{AI User}, \texttt{AI Model} \\
        \hline 
        Regulation (EU) No 1025/2012 on European standardisation~\cite{eu-standardisation-reg} & 1 & \texttt{Standard}\\
        \hline

    \end{tabularx}
    
    \label{tab:<ch4-airo-ontology-sources>}
\end{table*}

In addition to the iterations triggered by publication of standards, AIRO was revised throughout the AI Act's ordinary legislative process to incorporate the changes introduced in different phases. The extent of changes to AIRO's concepts and relations were minimal, primarily affecting the definitions extracted from Article 3 of the Act. 

It should be noted that in the process of ontology development, initially all the concepts were included in a single ontology. As the ontology expanded with population of the concepts, it became less reusable and more complex to manage. Therefore, the ontology was broken down into a minimal ontology (AIRO), that provides foundational concepts and relations, and a separate vocabulary (VAIR) extending AIRO with hierarchies of instances.

\subsection{AIRO Overview} \label{section:<ch4-ontology-airo>}

\begin{figure*} [!h]
  \centering
   \fbox{\includegraphics [width=0.8\textwidth]{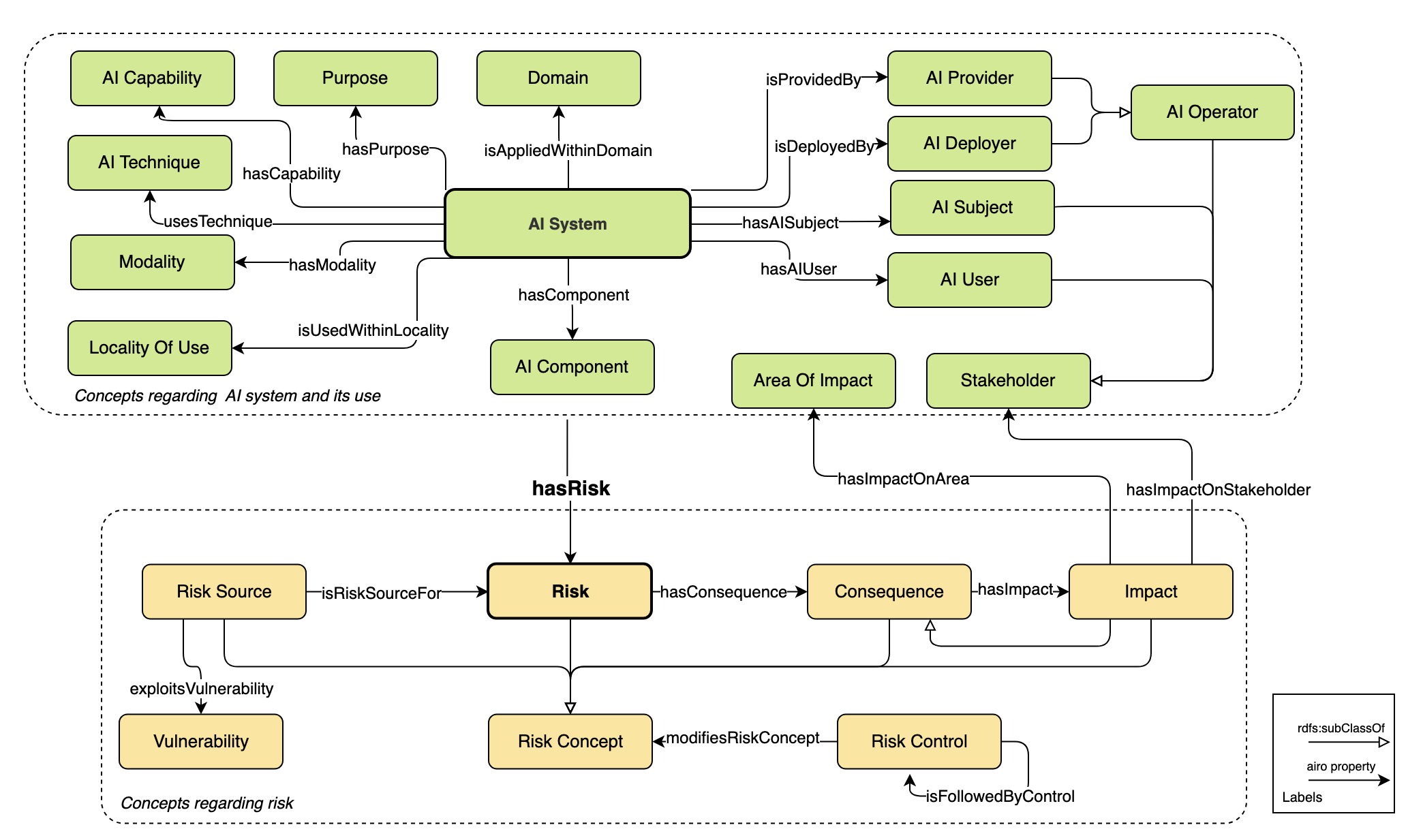}}
   \caption{Overview of AIRO's main concepts and relations}
   \label{fig:<ch4-airo>}
\end{figure*}

AIRO’s core concepts and relations are illustrated in \autoref{fig:<ch4-airo>}. The upper half shows the main concepts required for describing a specific use of an \texttt{AI System} (green boxes), and the lower half represents key concepts for expressing \texttt{Risk} (yellow boxes). The relation \texttt{hasRisk} links these two halves by connecting risk to an AI system or a component of the system. AIRO incorporates 56 classes, 61 object properties, and 3 data properties. 10 of the concepts, 9 of the object properties, and all of the data properties are reused from existing ontologies (see \autoref{ss:airo-reuse} for details).

The core concepts related to the use of an AI system are as follows:
\begin{itemize}
    \item \texttt{Domain}: indicates the area, sector, or industry in which the AI system is used.
    
    \item \texttt{Purpose}: specifies end goal for which the system is used.
    
    \item \texttt{Locality Of Use}: specifies the locality the system is designed to be used within, e.g. publicly accessible spaces. 
    
    \item \texttt{AI User}: refers to any individual or group that interacts with a system (ISO/IEC TR 24028, 3.43).
    
    \item \texttt{AI Subject}: represents an entity that is subject to or impacted by the use of AI.

\end{itemize}

Regarding technical details, the key concepts are: 
\begin{itemize}
    \item \texttt{AI Technique}: refers to computer science approaches and techniques used in development of a system, e.g. supervised machine learning. 
    \item \texttt{AI Capability}: refers to the capability of an AI system that enables realisation of the system's purposes, e.g. facial recognition.
    \item \texttt{AI Component}: specifies incorporating components of the system.  
    \item \texttt{Modality}: indicates the way or type in which the system exists, e.g. service. 
\end{itemize}
 
 The key actors involved in the AI development and deployment that are modelled in AIRO are as follows:

 \begin{itemize}
     \item \texttt{AI Provider}: is a natural or legal person, public authority, agency or other body that develops an AI system or a general-purpose AI model or that has an AI system or a general-purpose AI model developed and places it on the market or puts the AI system into service under its own name or trademark, whether for payment or free of charge (EU AI Act, Article 3(3)).
     
     \item \texttt{AI Deployer}: is any natural or legal person, public authority, agency or other body using an AI system under its authority except where the AI system is used in the course of a personal non-professional activity (EU AI Act, Article 3(4)).
     
     \item \texttt{AI Developer}: is an organisation or entity that is concerned with the development of AI services and products (ISO/IEC 22989, 5.19.3.2).
     
     \item \texttt{AI Operator}: refers to a provider, product manufacturer, deployer, authorised representative, importer or distributor (EU AI Act, Article 3(8)).
     
 \end{itemize}

Concerning risk, the key concepts are:
\begin{itemize}
    \item \texttt{Risk}: The AI Act defines risk as \emph{``the combination of the probability of an occurrence of harm and the severity of that harm''}. Risk in ISO 31073 is defined as \emph{``effect of uncertainty on objectives''}, with a note indicating that the effect can be negative, positive, or both. While both definitions provide quantitative perception of risk, ISO 31073 considers a broader scope of effects by including positive impacts, compared to the AI Act that only focuses on harms, i.e. negative effects. On another hand, the AI Act's definition of risk, together with the language used throughout its text, indicates a focus on harm to fundamental rights, health, and safety, whereas ISO focuses on broad organisational objectives. In AIRO, we  propose a novel definition of \texttt{Risk} as \textbf{the state of uncertainty associated with an AI system, that has the potential to cause harms and is expressed in terms of risk sources, consequences, impacts, likelihood, and severity}. To associate a specific risk to an AI system or a component of the system, we defined the \texttt{hasRisk} relation.
    
    \item \texttt{Vulnerability}: refers to properties of an entity, e.g. AI system or AI component, resulting in susceptibility to a risk source (ISO 31073:2022, 3.3.21 with modifications).
    
    \item \texttt{Risk Source}: indicates the element which alone or in combination has the potential to give rise to risk (ISO 31073:2022, 3.3.10). 
    \texttt{Threat} and \texttt{Hazard} are both types of risk sources that have potential to cause harm.
    
    \item \texttt{Consequence}: is the direct outcome of a risk that affects objectives (ISO 31073:2022, 3.3.18, event is replaced with risk for better accuracy). 
    
    \item \texttt{Impact}: represents the outcome of a consequence on individuals, groups, society, and environment. While consequence indicates the direct outcome of a risk on objectives, which may or may not involve external entities, 
    the notion of impact puts emphasis on the effect and influence on individuals, groups, and society. This distinction addresses the concern raised by Soler et al.~\cite{soler2023analysis} in regard to non-alignment of ISO/IEC and AI Act, rooted in organisation-focused view of risks in ISO/IEC documents that misses the AI Act's centre of attention: the effect of AI on individuals' health, safety, and fundamental rights.  
    
    \item \texttt{Risk Control}: ISO 31073 defines \emph{risk control} as a measure that maintains and/or modifies risks\footnote{ISO 31073:2022, 3.3.33}. In AIRO, risk control is not restricted to risks and indicates a measure that is applied to detect, mitigate, or eliminate risk sources, risks, consequences, or impacts. %Therefore, \texttt{Control} refers to any measure that detects, mitigates, or eliminates an \emph{event}.
\end{itemize}

To enhance flexibility in modelling various AI use cases, domain or range of some properties are not restricted to any concepts, for example for \texttt{airo:hasRisk}, no domains is specified. Another design decision made concerning object properties in AIRO is the exclusion of inverse relations to avoid redundancy and maintain simplicity.
%- the reasoning and querying do not need inverse relations to be defined explicitly. }
AIRO is published online at \url{https://w3id.org/airo}\footnote{Also available on Zenodo at \url{https://doi.org/10.5281/zenodo.10894952}} under the CC-BY-4.0 license and is indexed in the linked open vocabularies (LOV) registry at \url{https://lov.linkeddata.es/dataset/lov/vocabs/airo}. %Evaluation of AIRO will be discussed in \autoref{section:<ch7-airo-vair-evaluation>}. 

\subsection{Ontology Reuse} \label{ss:airo-reuse}
%As discussed in the previous chapter, the ontologies created in this work require to support generation of documentation. Therefore, the competency questions outlined in \ref{section:<ch4-aicards-requirments-for-ontology>} served as ontology requirements and guided the development and revision of AIRO. 

In addressing competency questions, as per established Semantic Web practices, where more specialised and well-known open ontologies exists, appropriate concepts and relations are reused from them, rather than modelling them from scratch. In AIRO, classes and properties from Data Protection Vocabulary (DPV)\footnote{\url{https://w3id.org/dpv/}}~\cite{pandit2024dpv}, Data Quality Vocabulary (DQV)\footnote{\url{	http://www.w3.org/ns/dqv}}~\cite{albertoni2021dqv}, Data Catalog Vocabulary (DCAT)~\cite{dcat-v3}, and DCMI Metadata Terms~\cite{dcterms} are reused. % for describing information related to data processing and quality. 
The list of reused concepts and properties is presented in \autoref{tab:<ch4-airo-reuse>}. 

  %There is a lack of FAIR ontologies for AI risks and \att{AIRO is the first formal ontology available for expressing risks of AI systems according to the AI Act}. 

\begin{table*}[!h]
    \centering
    \small
    \caption{Ontologies reused in AIRO}
    \begin{tabularx}{0.8\textwidth}{|p{1.4cm}|X|X|p{3cm}|}
    \hline
    \textbf{Source} & \textbf{Classes} & \textbf{Object properties} & \textbf{Data properties}\\
        \hline
   
       \texttt{\textbf{DCAT}} & \texttt{dcat:Dataset} & None & None  \\

      \hline
        \texttt{\textbf{DPV}} & \texttt{dpv:Processing},  \texttt{dpv:Data}, \texttt{dpv:SensitiveData},  \texttt{dpv:DataSource}, \texttt{dpv:DataSubject}, \texttt{dpv:LegalBasis} & \texttt{dpv:hasProcessing},  \texttt{dpv:hasData}, \texttt{dpv:hasDataSource},  \texttt{dpv:hasDataSubject},  \texttt{dpv:hasLegalBasis} & None \\
        
        \hline
        \texttt{\textbf{DQV}} &  \texttt{dqv:QualityMeasurement},  \texttt{dqv:Metric}, \texttt{dqv:Dimension} &  \texttt{dqv:hasQualityMeasurement}, 
        \texttt{dqv:isMeasurementOf}, 
        \texttt{dqv:inDimension}, 
        \texttt{dqv:expectedDataType}& \texttt{dqv:value} \\
        \hline
        \texttt{\textbf{DCT}} & None & None & \texttt{dct:date}, \newline
        \texttt{dct:title} \\        \hline
    \end{tabularx}
    \label{tab:<ch4-airo-reuse>}
\end{table*}

\section{VAIR: A Vocabulary of AI Risks} \label{section:<ch4-vair>}

The high-level model provided by AIRO is not sufficient for annotating AI use cases, expressing rules for identification of the AI Act's high-risk AI systems, and documenting and registering AI and risk information. These require enriching AIRO with specialisations of concepts that are represented formally and organised in hierarchies. The Vocabulary of AI Risks (VAIR) is a specialisation of AIRO, providing a taxonomy for expressing and exchanging information regarding AI systems and their risks aligned with the requirements of the AI Act. 
 
\subsection{Iterative Development of VAIR} \label{section:<ch4-vair-iterations>}
 
 %In the first iteration
 As previously mentioned, the first version of AIRO included key concepts plus their specialisations, which were modelled as sub-classes. However, emergence of new standards and publication of multiple AI Act mandates have shown that while the key top-level concepts remain largely intact, new sub-classes are likely to be iteratively identified from multiple sources as the understanding of AI risks evolve over time. %This implied a pattern that when unknown became known, the extent of changes to AIRO were  minimal. %T
 
 Therefore, the instances were de-integrated from AIRO and provided under the VAIR namespace. In this modular structure, AIRO acts as a foundational ontology, including only key high-level concepts and relations, while VAIR serves as a specialisation of AIRO that contains a taxonomy of these key concepts. This design decision contributes to better reusability of AIRO and VAIR, in particular for creating domain-specific risk taxonomies.

%and further enhances expansion of VAIR 
 %This implies a pattern  that when unknown became known, AIRO was not affected that much %(as mentioned earlier the wording in the definitions were the most common amendment) but VAIR requires mainly addition of concepts. Further, 
 
 VAIR was updated iteratively, upon publication of different mandates of the AI Act. In addition to the AI Act, the EU AI Watch's\footnote{AI Watch is the European Commission's JRC knowledge service on AI, that supports development and uptake of trustworthy AI in the EU.} AI taxonomy~\cite{jrc2020aiwatchtaxonomy, jrc2021aiwatchtaxonomy2} and ISO/IEC 22989:2022 on AI terminology~\cite{iso22989} were used to further populate VAIR.  
 
 Providing an exhaustive and commonly-agreed taxonomy in this work was not practical, firstly, due to the extensive state of the art and non-Semantic Web efforts to build AI risk taxonomies. Secondly, building such a taxonomy requires involvement of stakeholders to build consensus around the taxonomy. Therefore, VAIR focuses on inclusion of the concepts that are essential to satisfy the competency questions and are needed for implementing the Semantic Web approaches that are in the scope of this work. After the publication of the final version of the AI Act in 2024, certain concepts previously extracted from different mandates of the Act were no longer \textit{essential} for addressing the competency question. For example, changes to the list of high-risk AI applications in Annex III led to removal of some \texttt{Purpose}s, however, these concepts were intentionally retained in VAIR for better comprehensiveness.  
 
Regarding stakeholders' involvement in the development process, as will be discussed in \autoref{s:dpv_integration}, both AIRO and VAIR were proposed to the W3C Data Privacy Vocabularies and Controls Community Group\footnote{\url{https://www.w3.org/groups/cg/dpvcg/}} (DPVCG) and has been included in DPV through a consensus-based process.

\section{Overview of VAIR} \label{s:vair}
VAIR is a hierarchical taxonomy, providing a specialisation of AIRO's concepts and therefore serves as a vocabulary for expressing AI use cases in an interoperable manner. VAIR is provided as an OWL 2 ontology where the concepts are defined using both \texttt{skos:Concept} and \texttt{rdfs:Class}, with hierarchical relations between them modelled using \texttt{rdfs:subClassOf}. This promotes extensibility and reusability by enabling the concepts to be used as both instances and classes in specification of an AI system.
%is used to define the hierarchies.  %Following the DPV's style of modelling \cite{pandit2024dpv}, the VAIR's data model also provided using RDFS+SKOS serialisation, where the concepts are defined using both \texttt{skos:concept} and \texttt{rdfs:Class} with hierarchical relations between them modelled using \texttt{skos:broader}. This promotes extensibility and reusability by enabling the concepts to be used as both instances and classes in specification of an AI system.}

 %utilises the SKOS (Simple Knowledge Organization System) model\footnote{\url{https://www.w3.org/TR/2009/REC-skos-reference-20090818/}} to model concepts 

In total, VAIR includes 402 classes, with the following thematically-categorised taxonomies:  
\begin{itemize}
    \item \textbf{AI}: contains taxonomies of \emph{techniques} (number of instances in the taxonomy: 17), \emph{capabilities} (32),  \emph{types} of AI systems (13), \emph{components} (23), \emph{lifecycle phases}~(10), %{characteristics} (20) including trustworthiness characteristics, 
    \emph{modality} (4) and \emph{outputs} (5).
    
    \item \textbf{Use of AI}: includes taxonomies for defining AI use cases namely \emph{purposes} (118) and \emph{domains} (11). In addition, 7 \emph{levels of automation}, 6 types of \emph{human involvement}, and 6 \emph{modes of output controllability} are included.
    \item \textbf{Risk}: contains taxonomies of \emph{risk sources} (44), \emph{consequences} (4), \emph{impacts} (5), \emph{risk controls} (20), and \emph{areas of impact} (7).
    \item \textbf{Stakeholder}: contains types of \emph{AI operators} (17) and \textit{AI subjects} (17), mainly extracted from Annex III of the AI Act. 
    \item \textbf{Document and standard}: contains a list of  \emph{documentation} (13), including those required for conformity assessments, and \emph{standards} (23) that can potentially be used in implementation of the AI Act.
\end{itemize}

 VAIR is published online as an open resource under the CC-BY-4.0 license at \url{https://w3id.org/vair} and is indexed in LOV at \url{https://lov.linkeddata.es/dataset/lov/vocabs/vair}\footnote{Also available on Zenodo at \url{https://doi.org/10.5281/zenodo.10894914}}.

\section{Evaluation}
\subsection{Demonstartion of Applicabilty through Proof Of Concept} \label{ss:poc}

In this section, we demonstrate the applicability of AIRO and VAIR in modelling AI use cases and their associated risks. To this end, one synthetic use case was created and two incidents from the AIAAIC were selected, based on the following criteria:
\begin{itemize}
    \item The initial analysis of the use case demonstrates that it is highly likely to be  classified as high-risk as per Annex III, 
    \item The domains in which the systems are used should not overlap. This is to demonstrate the capability of the ontologies to be used in and/or extended for various domains,
    \item Diversity in types of  harmful impacts is preferred in order to include harms to health, safety, and fundamental rights. 
    \item For use cases selected from the AIAAIC repository, documentation that provides detailed information about the AI system and its risks should be available. In this, topicality of the use case is preferred. 
\end{itemize}

Based on these criteria, the following use cases were selected to be modelled:
\begin{itemize}
    \item \textbf{Use case \#1: Proctify}: is a synthetic, but realistic, AI-based student proctoring tool, inspired by the use cases described in \cite{panigutti2023role} and \cite{hupont2023usecasecards}.  It is a potentially high-risk AI use case in the education domain as per Annex III, Point 3(d). This use case has been designed to show the potential of AIRO and VAIR, without being restricted to model the information provided by third-parties. 
    
    \item \textbf{Use case \#2: Uber’s Real Time ID (RTID)} (incident ID = AIAAIC0756): is a system to verify the identity of Uber drivers using facial recognition. This incident was selected as an example of AI systems used in the employment domain (Annex III, Point 4). The system has negatively affected certain demographics and has led to violation of fundamental rights.
    
    \item \textbf{Use case \#3: The Spanish Ministry of the Interior's AI system (VioGén)} (incident ID = AIAAIC0848): is an AI system that is being used for assessing risk of domestic violence. This use case is related to both public services (Annex III, Point 4) and law enforcement (Annex III, Point 6) domains. The incidents caused by VioGén have already impacted health and safety of victims of gender violence.   
\end{itemize}

In the modelling process, each use case was initially annotated manually using AIRO. If the identified individual, which was annotated as an \texttt{rdf:type} of an AIRO concept, belonged to a more specific class in VAIR, then it was defined as \texttt{rdf:type} of that specific VAIR class. If the individual matched an \texttt{owl:NamedIndividual} in VAIR, it was replaced with the corresponding VAIR concept.
It should be emphasised that our aim here is to show applicability of AIRO and VAIR in modelling use cases, rather than illustrating their comprehensiveness, given that development of comprehensive AI risk ontologies was not feasible.

In the following we provide short description of each use case, alongside their machine-readable specifications using AIRO and VAIR.

\subsubsection{Use Case \#1: Proctify}
Proctify is designed to detect suspicious behaviour during online exams by analysing facial behaviour from a student's facial video captured throughout the exam using a webcam. The system incorporates a graphic interface displaying an analysis of the student's face including the head pose, gaze direction, and face landmarks' positions. This extracted information is then provided as an input to \emph{SusBehavedModel}, which has been trained in-house by the system's provider using \emph{SusBehavedDataset}, to determine whether the student is displaying suspicious behaviour, e.g. looking away from the screen, leaving the room, or a third person detected in the room. Detection of suspicious behaviour raises an alarm in the interface to inform and let the human oversight actors, e.g. human instructors, take appropriate actions, e.g. communicating with the student.

A snippet of the machine-
readable specification is provided in Listing \ref{lst:<proctify>}. The comprehensive  machine-readable representation of Proctify using AIRO and VAIR in Turtle serialisation is available online on GitHub\footnote{\url{https://github.com/DelaramGlp/airo/blob/main/usecase/proctify.ttl}}.

\begin{lstlisting}[caption={A snippet of machine-readable provision of Proctify in Turtle}, label ={lst:<proctify>}]
@prefix ex: <https://example.com/proctify#> .
@prefix airo: <https://w3id.org/airo#> .
@prefix vair: <https://w3id.org/vair#> .
@prefix rdf: <http://www.w3.org/1999/02/22-rdf-syntax-ns#> .
@prefix dcterms: <http://purl.org/dc/terms#> .
@prefix xsd: <http://www.w3.org/2001/XMLSchema#> .
@prefix dqv: <http://www.w3.org/ns/dqv#> .
ex:proctify
    a airo:AISystem ;
    airo:hasVersion ex:v_1.0.2 ;
    dcterms:date "2023-09-11"^^xsd:date ;
    airo:hasModality vair:Software ;
    airo:usesTechnique vair:DeepLearning ;
    airo:hasInput ex:facial_video ;
    airo:producesOutput ex:suspicious_behaviour_alarm ;
    airo:isProvidedBy ex:AIEduX ;
    airo:isDevelopedBy ex:AIEduX ;
    airo:hasComponent ex:facial_analysis_toolkit,
                      ex:susbehaved_dataset ;
    airo:hasModel     ex:susbehaved_model ;             
    airo:isAppliedWithinDomain vair:Education ;
    airo:hasPurpose vair:DetectingProhibitedBehaviourDuringTest,
                    ex:facial_behaviour_analysis,
                    ex:video_analysis ;
    airo:hasCapability ex:facial_recognition ;
    airo:isDeployedBy ex:university ;
    airo:hasAutomationLevel vair:PartialAutomation ;
    airo:hasAISubject vair:Student,
                      ex:other_occupant ;
    airo:hasAIUser ex:instructor ;
    airo:hasRisk ex:inaccuracy_risk_for_darker_skin ;
    airo:hasRiskControl ex:bias_testing ;
    dqv:hasQualityMeasurement ex:alarm_precision_measurement,
                              ex:alarm_recall_measurement, 
                              ex:alarm_f_score_measurement   ;
    airo:hasPreDeterminedChange ex:change_of_model ;
    airo:compliesWithRegulation ex:EU_GDPR ,
                                ex:Irish_Data_Protection_Act ;
    airo:conformsToStandard vair:ISOIEC42001_2023 ,
                            vair:ISOIEC27001_2022  .                        
    
    

\end{lstlisting}

\subsubsection{Use Case \#2: Uber's Real-time ID Check System}

\begin{figure*} [ht]
  \centering
   \fbox{\includegraphics [width=0.8\textwidth]{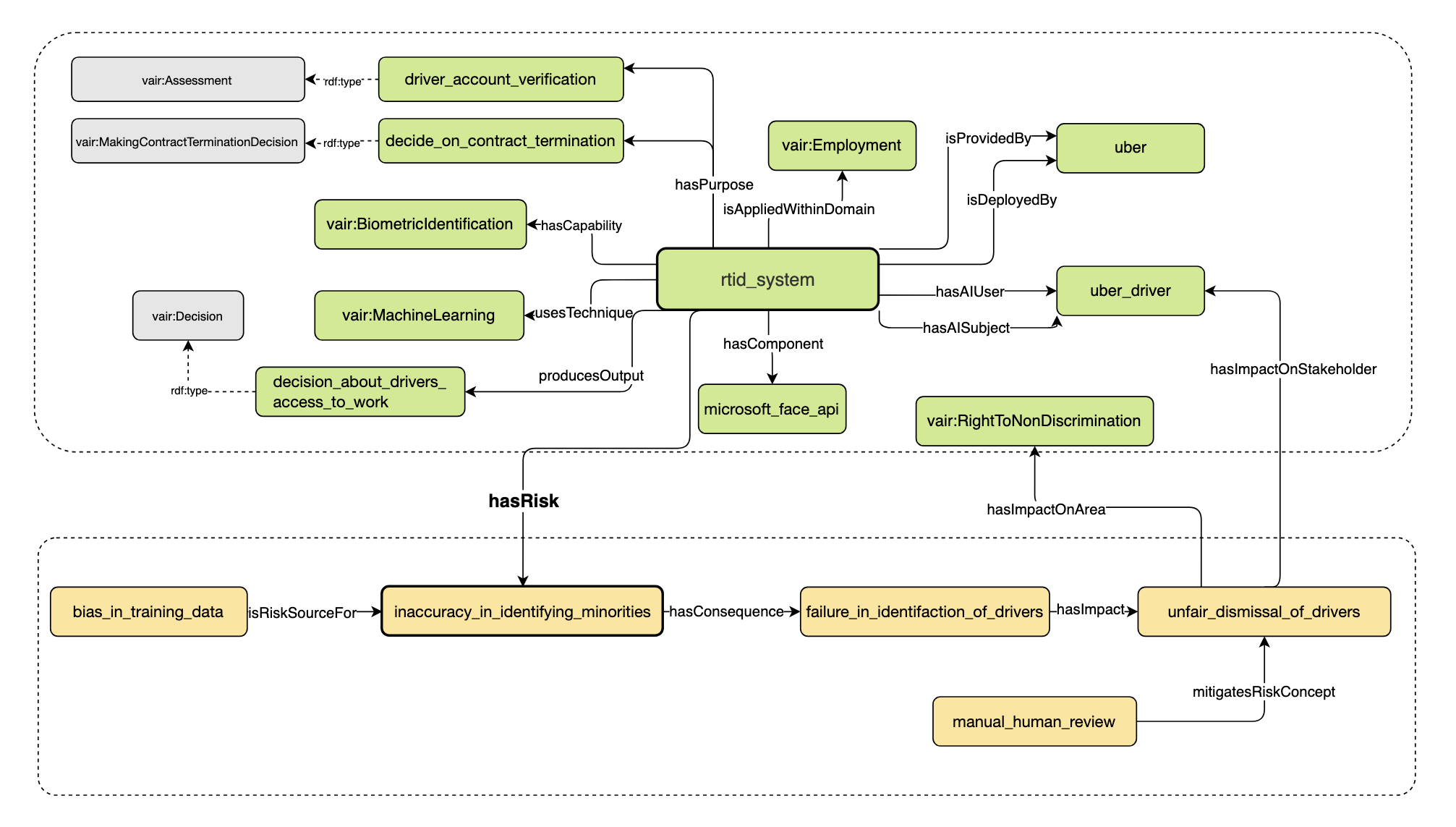}}
   \caption{Visual representation of the graph specifying the Uber's Real-time ID Check use case}
   \label{fig:<ch4-uber>}
\end{figure*}

Uber's facial recognition identification system, known as the Real Time ID (RTID), aimed to ensure that the driver's account is not used by anyone other than the registered Uber driver. When the system failed to recognise a person for two consecutive times, the driver's contract would be terminated and their driver and vehicle licenses would be revoked. Multiple incidents were reported concerning the system's failure in verifying drivers, in particular those belong to ethnic minorities or vulnerable groups, leading to unfair dismissal of drivers (vide \cite{melendez2017uber} and \cite{vallancez2021uber}). This system can be considered high-risk as per Annex III, Point 4(b) as it is used to ``make decisions affecting terms of work-related relationships''. \autoref{fig:<ch4-uber>} depicts the graph expressing Uber's Real-time ID Check use case using AIRO and VAIR, with the machine-readable specification available online on GitHub\footnote{\url{https://github.com/DelaramGlp/airo/blob/main/usecase/uber.ttl}}.

\subsection{Use Case \#3: VioGén Domestic Violence Risk Assessment System}
\begin{figure*} [!h]
  \centering
   \fbox{\includegraphics [width=0.8\textwidth]{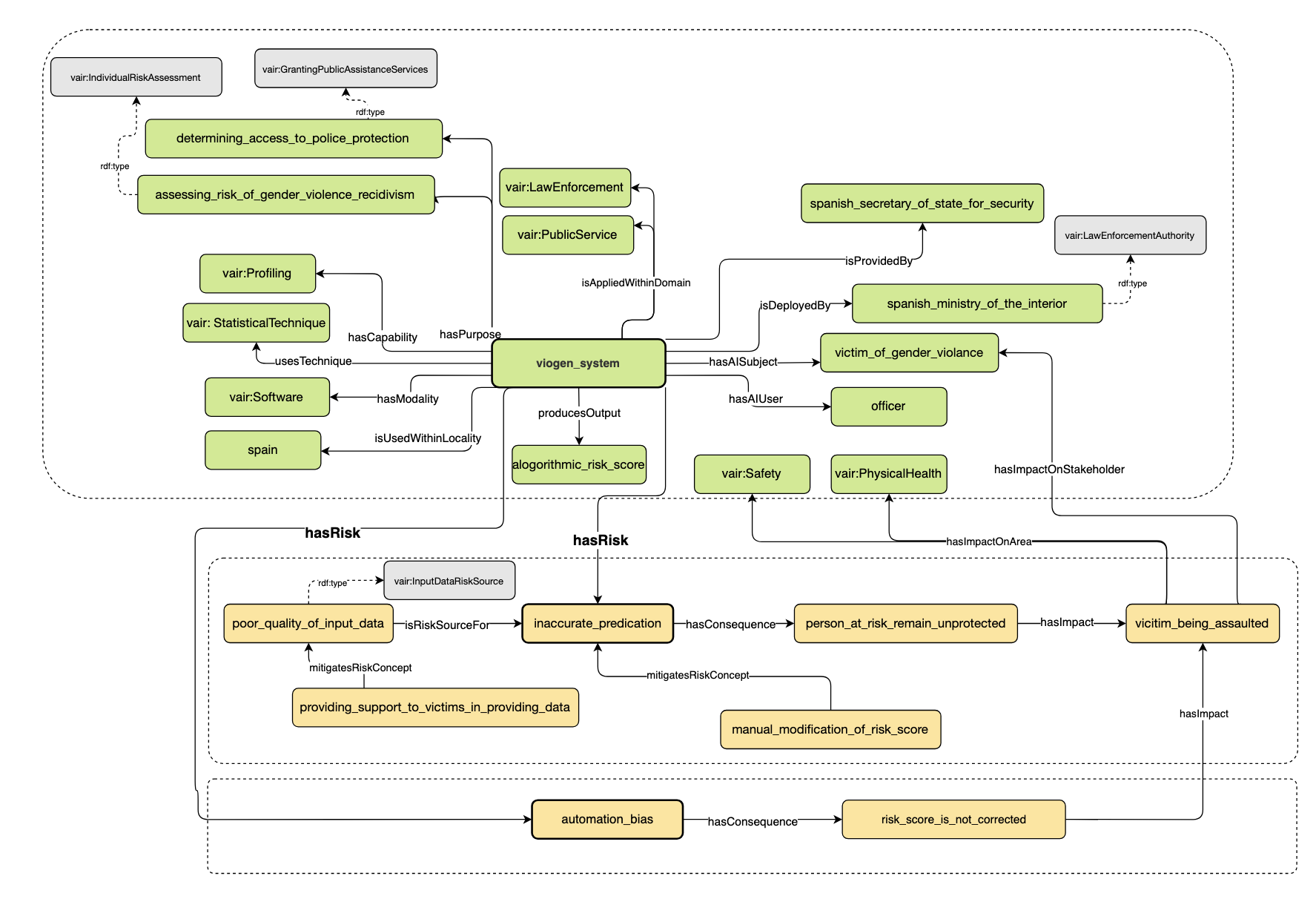}}
   \caption{Visual representation of the graph specifying the VioG{\'e}n use case}
   \label{fig:<ch4-viogen>}
\end{figure*}

%https://www.nytimes.com/interactive/2024/07/18/technology/spain-domestic-violence-viogen-algorithm.html
 VioGén System is used by the Spanish law enforcement agencies to determine the victim's eligibility for police protection by assessing the likelihood of a victim of gender violence to be assaulted by the same perpetrator again~\cite{alvarez2018viogen}. Its use of statistical models to predict the risk faced by a victim raised questions regarding the accuracy of its predictions since these would be highly dependent on the quality of data fed into the models. The input data was generated based on a questionnaire answered by victims who filed a report. The ambiguity of questions and timing of questionnaire could have lead to inaccurate or biased predictions, and if the score was not modified by police officers, the victims would not receive the police protection they need. To address this risk, police officers were granted the power to increase the risk score calculated by the system. However, according to \cite{eticas2022viogen}, in most cases the officers trusted the system's scoring despite warning signs, which led to ``automation bias'', i.e. over-reliance on the system's outcomes. Under the AI Act, VioGén is likely to be high-risk as it is used by the police to assess the ``risk of a natural person becoming the victim of criminal offences'' (Annex III, Point 6(a)). It also can fall under the high-risk AI systems used by public authorities to ``evaluate the eligibility of natural persons for essential public assistance'' (Annex III, Point 5(a)), in this case police protection. The graph expressing the VioGén use case is provided online on GitHub\footnote{\url{https://github.com/DelaramGlp/airo/blob/main/usecase/viogen.ttl}} and is illustrated in \autoref{fig:<ch4-viogen>}. 

\subsection{Demonstration of Applicability thorough Semantic Web-based Applications}
AIRO and VAIR are both used in implementation of a set of approaches to assist with AI Act compliance tasks utilising different technologies from the stack of Semantic Web technologies and existing W3C standards. As described in detail in \cite{golpayegani_clsr}, these approaches include:
\begin{enumerate}
    \item  A set of SHACL shapes~\cite{shacl} for determining Article 5 prohibited and Annex III high-risk AI systems as per Annex III, 

    \item  AIUP (AI Use Policy) profile that extends the Open Digital Rights Language (ODRL)~\cite{odrl} to describe intended, precluded, and conditions of AI use as described by the AI Act, 
    
    \item A set of SPARQL~\cite{sparql} queries to generate risk management documentation and technical documentation according to Annex IV, 
    
    \item AICat as an extension of DCAT~\cite{dcat-v3} for cataloguing AI systems and their components in AI registries including the  the EU database of high-risk AI systems.
\end{enumerate}

\begin{figure*}
    \centering
    \fbox{\includegraphics[width=0.6\linewidth]{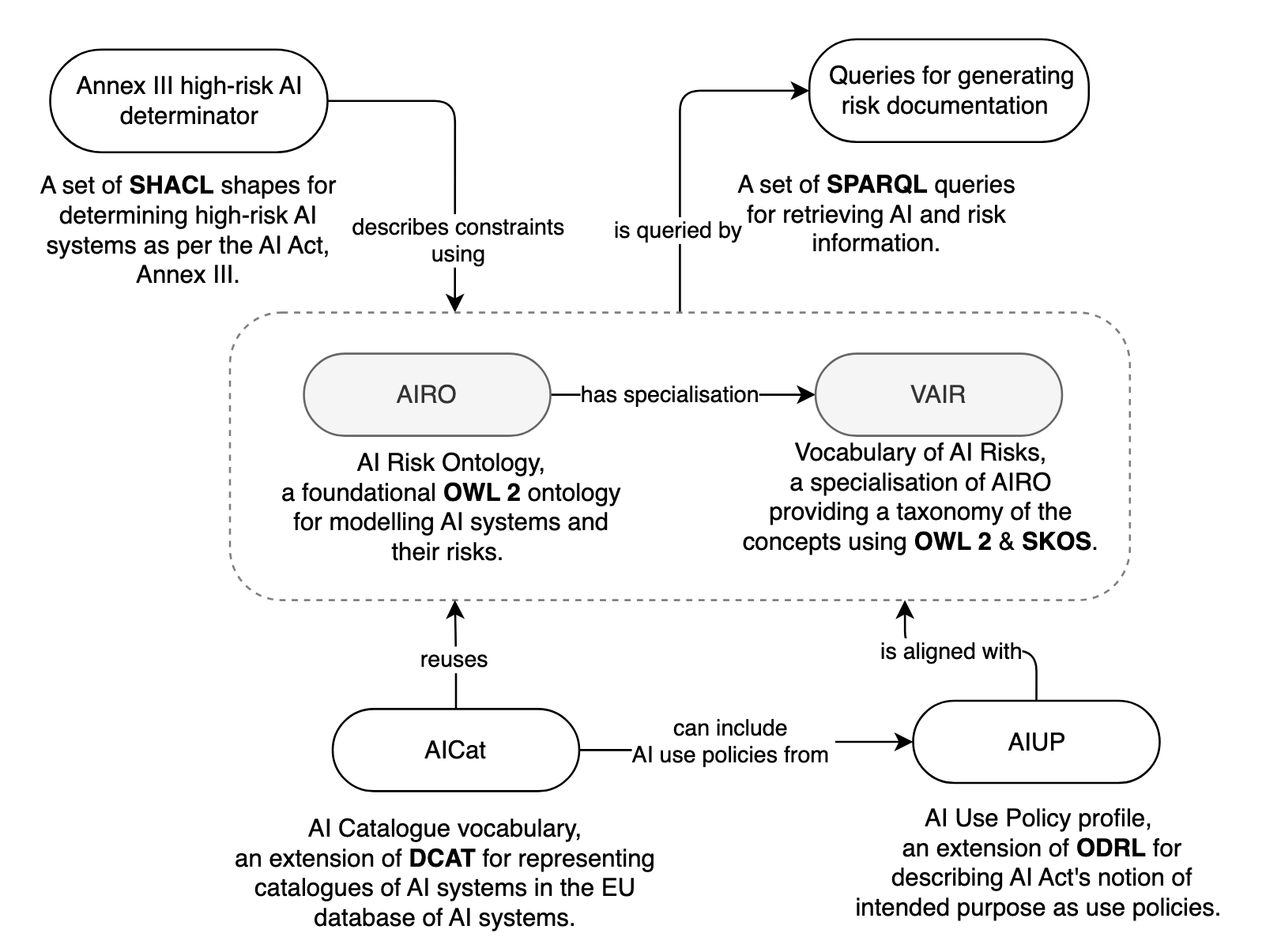}}
    \caption{The set of Semantic Web-based artefacts built upon AIRO and VAIR, as detailed in~\cite{golpayegani_clsr} }
    \label{fig:<artefacts>}
\end{figure*}

Although this paper does not apply AIRO and VAIR  in real world settings, they have already been adopted in several academic and industry projects (see \autoref{s:impact}), demonstrating their applicability and usability in practice.

\subsection{Quality Assessment}\label{ss:eval-quality}
Quality assessment of AIRO and VAIR was carried out in accordance with the following criteria:

\begin{enumerate}
    \item \textbf{Syntax validity and logical consistency}: Throughout the development process, the syntax was checked using Protégé. Logical consistency was tested using build-in Protégé reasoners including FaCT++~1.6.5 and HerMiT~1.4.3. 
    
    \item \textbf{Adherence to ontology development best practices}: In addition to following LOT as an ontology development methodology, quality of AIRO and VAIR was ensured by following best practices and guidelines from the Semantic Web community, including 
    the W3C's Data on the Web Best Practices~\cite{w3c_dataonweb}, the W3C Recipes for Publishing RDF Vocabularies~\cite{rdf-bespractice}, FAIR best practices for vocabularies and ontologies~\cite{garijo2020fair}, and WIDOCO best practices for ontology documentation~\cite{garijo2017widoco}. 
    
    \item \textbf{Freedom from common ontology pitfalls}: OOPS! (OntOlogy Pitfall Scanner!)~\cite{poveda2014oops}, was used to detect and further resolve common errors and pitfalls of AIRO and VAIR. The key unresolved pitfalls raised by OOPS! concerned the following:
    \begin{itemize}
        \item Missing annotations: It was raised due to the absence of definitions for those classes and properties reused from other ontological resources.
      \item Missing disjointness: This issue was raised as disjointness between some classes are not defined, given that they can share individuals. %For instance, a particular {airo:Consequence} and  become {airo:RiskSource}  for another risk.   
      
        \item Missing domain or range in properties: This pitfall addressed the deliberate design decision, discussed in \autoref{section:<ch4-ontology-airo>}, to enhance AIRO's flexibility in modelling AI use cases by not restricting domain or range for some properties. 
        \item Inverse relationships not explicitly declared: This issue addressed the lack of  inverse relations (\texttt{owl:inverseOf}) between object properties. This was intentional to keep AIRO minimal and simple.  
        \item Equivalent classes not explicitly declared: This was a false positive, incorrectly identifying equivalent classes. %within AIRO.
    \end{itemize}
    %\att{Regarding VAIR, the following pitfalls were identified by OOPS!}
    
    \item \textbf{FAIRness}: Adherence of the ontologies to FAIRness principles was evaluated using FOOPS! \cite{garijo2021foops}. The fairness scores for both AIRO and VAIR, as calculated by the FOOPS! web service\footnote{\url{https://foops.linkeddata.es/FAIR_validator.html}}, showed their full compliance with FAIRness principles for ontologies. 

\end{enumerate}

\section{Integration of AIRO and VAIR with DPV} \label{s:dpv_integration}

At the time of developing AIRO and VAIR, DPV~\cite{pandit2019dpv1,pandit2024dpv} represented the state-of-the-art in machine-readable vocabularies for representing personal data processing based primarily on the GDPR, and its use to develop specifications for tasks such as consent management~\cite{pandit_2024_iso_consent} and maintaining records of processing activities~\cite{ryan2022dpcat}. This reflected the scope through which the DPVCG was established in 2019, and the consensus of its members in developing DPV over the years. When developing AIRO and VAIR, the authors identified several overlaps in concepts, such as data, which have been discussed earlier in this paper. Based on this, the authors opened discussions within the DPVCG regarding its scope and how to address 'AI' as a technology and its regulation. After deliberations, the group agreed to expand its scope to include AI as a form of technology, and to restructure the DPV into several modular groups or `extensions', such as to represent AI and laws such as GDPR, AI Act \cite{pandit2024dpv}. Following this, concepts from AIRO and VAIR were proposed and accepted towards creating new extensions and enriching existing ones as follows:
\begin{itemize}
    \item  The EU AI Act extension (\texttt{EU-AIACT})\footnote{ available at \url{https://w3id.org/dpv/legal/eu/aiact}} was created;
    
    \item  The AI extension (\texttt{AI})\footnote{\url{https://w3id.org/dpv/ai}} was created;
    
    \item The Technology extension (\texttt{TECH}) \footnote{available at \url{https://w3id.org/dpv/tech}} was expanded;
    
    \item The Risk extension (\texttt{RISK})\footnote{available at \url{https://w3id.org/dpv/risk}} was expanded.

\end{itemize}

The integration of AIRO and VAIR represented the first instance of an `major externally developed resource' being integrated into the DPV, as prior to this the group worked on specific concepts based on strictly scoped taxonomies such as data categories. It also represented the first instance of developing additional vocabularies for a new law (AI Act) while trying to maintain the legal relevance (GDPR). In turn, the integration and DPVCG's consensus led process provided AIRO and VAIR an additional measure of ontology assessment as the group had to accept not only the concepts but also their relevance within the larger DPV-based vocabularies. The DPVCG also provides discoverability and findability as the DPV already has strong adoption, including use in standards, and a cycle for continuous maintenance based on stakeholder engagement. More importantly, it lays the ground for development of RegTech solutions that facilitate tasks related to compliance with multiple AI and data regulations using a single vocabulary.

Integration into DPV does not preclude the independent use of AIRO and VAIR. As discussed in the following section, these ontologies are valuable in their own right, since their narrower, single-purpose scope makes them simpler to reuse.

\section{Impacts and Uptake of AIRO and VAIR} \label{s:impact}
% published under permissive licences that encourages reuse, subject to attribution.
This research has been continually progressed with the AI Act's development process, until it was officially accepted as an EU law and published in the Official Journal of the European Union in July 2024. Consequently, the results of this work have increasingly attracted interest from the academics and wider audiences. To identify the impacts of AIRO and VAIR, we analysed the studies citing our prior publications on these two ontologies, namely~\cite{golpayegani2022airo} and \cite{golpayegani2023high} indexed by Google Scholar, as of July 2026. After excluding first-author self-citations, non-English publications, duplicates, and papers without accessible full text, a total of 121 studies, including  papers (both peer-reviewed and non-peer-reviewed) and theses  were analysed: 63 citing AIRO and 58 citing VAIR, with 10 appearing in both sets. The analysis aimed to determine why and how AIRO or VAIR were used. Our analysis identified seven categories of citation and reuse:

\begin{enumerate}
    \item \textbf{Background information}: our papers were cited to support statements related to the AI Act, AI risks, AI incidents, or relevant standards. Citations in this category typically referred to our work as a supporting reference for a minor legal or factual point and do not engage with our core contributions.
    
    \item \textbf{Related work}: our papers were cited as state-of-the-art ontologies, semantic frameworks, or approaches to risk management or AI Act compliance.
    
    \item \textbf{Reuse of ontologies}: AIRO, VAIR, or both were reused as Semantic Web artefacts, whether imported directly as machine-readable vocabularies or drawn upon as human-readable concept catalogues.  

    \item \textbf{Reuse of legal analysis}: our interpretation of the AI Act was used in the citing study. In particular, the five-concept structure for determining Annex III high-risk AI systems (domain, purpose, capability, AI user, AI subject) has be received very-well by the community. We categorised these studies into legal analysis categories, however they can also be considered as an ontology reuse at a conceptual level.   %for example as the basis of a risk-classification tool or assessment method. %The 5-concept framework for determining Annex III high-risk AI systems, proposed in~\cite{golpayegani2023high} has be received very-well by the community, we categorised them into legal analysis categories, however they can be considered as an ontology reuse at a conceptual level.  
    
    \item \textbf{Adoption of the ontology development approach}: our iterative and modular methodology for designing ontologies under uncertainty was adopted to construct  ontologies.

    \item \textbf{Evaluation of contributions}: the citing study evaluates the ontologies or validates our legal analysis. 
    
    \item \textbf{Planned future use}: the citing study stated an intention to adopt or integrate AIRO, VAIR, or both in the future work.
\end{enumerate}

\autoref{tab:<citation_overview>} summarises the distribution of citing works across theses seven citation categories. Note that a single work may fall into more than one category where it draws on our contributions in multiple ways, for example both mentioning AIRO as related work and reusing it within their ontology. As the table shows, \textit{related work} constitutes the largest share of citations, while \textit{methodology adoption} was only addressed by one contribution. In the following, we take a closer look into some of the prominent studies that reused our work. %legal analysis (category 3),  reused the ontologies (category 4), and adopt the methodology (category 5). 

\begin{table*}[h]
    \centering
    \caption{An overview of citation to AIRO and VAIR}
    \begin{tabularx}{\textwidth}{|X|X|X|X|X|X|X|X|X|}
    \hline
       Paper  & No. of works analysed &  1. Background information & 2. Related work & 3. Legal analysis reuse & 4. Ontology reuse & 5. Methodology adoption & 6. Evaluation of contributions & 7. Future work \\
       \hline
     AIRO~\cite{golpayegani2022airo} & 63 &     
     %(B)
     9 & 
     %(R)
     36 & 
     %(LR)
     3 & 
     %(OR) 
     22 & 
     %(OM)
     1& 
     % (Ev)
     0 &
     %(F)
     3\\
  
    \hline
     VAIR~\cite{golpayegani2023high} & 58 &  
     %(B)
     21 & 
     %(R)
     23 & 
     %(LR) 
     14 & 
     %(OR)
     8 & 
     %(OM) 
     1 &
     %(EV)
     2 & 
     %(F)
     1  \\
     \hline
    \end{tabularx}
    
    \label{tab:<citation_overview>}
\end{table*}

AIRO and VAIR have been reused across a range of applications. \textbf{Ontology reuse} concentrates on AIRO, as it provides the minimal core semantic model. AIRO has been reused for policy and regulatory compliance checking and reasoning~\cite{baldwin2026knowledgegraphrepresentationsllmbased, co2, goncalves2026operationally}, AI risk cataloguing~\cite{IBM_2026}, (probabilistic) AI risk analysis~\cite{Fumagalli2026news}, AI risk management~\cite{Kubesch2026exploring, IBM_2026,Daly2025usge_ibm, ikhsan2025pattern, bagehorn2025airiskatlastaxonomy}, %including risk mitigation and risk assessment
 fundamental rights impact assessment~\cite{rintamaki2024developing},  AI documentation~\cite{neumaier2026aidoc}, and bias documentation~\cite{russo2024biaso}. VAIR has been reused for AI documentation~\cite{neumaier2026aidoc}, bias documentation~\cite{russo2024biaso}, AI risk assessment~\cite{ikhsan2025pattern},  analysis of AI incidents~\cite{velazquez2024decoding}, and annotating AI flaw reports~\cite{longpre2025position}.

AIRO has also been taken up in specific domains. It has been reused as an information model within a preposed quality management methodology for AI in  medicine~\cite{elia2025generic}, in an ontology for managing linguistic uncertainty~\cite{csonka2025linguistic}, in an ontology for applied ethics~\cite{aijaz2025appleappliedethicsontology}, in  a risk schema for an AI-native enterprise architecture~\cite{gaikwad2026aegis}. 
Reuse of AIRO through DPV is not taken up yet and to date, we have identified two studies that reuses AIRO indirectly through DPV~\cite{rintamaki2024developing, clavell2026eticasairisktaxonomy}. 

%It is worth noting that our subsequent work demonstrates how AIRO and VAIR can be applied to different AI Act compliance tasks~\cite{golpayegani_clsr}, including determining high-risk and prohibited AI systems, cataloguing AI systems, creating and maintaining technical documentation~\cite{golpayegani2024AIcards}, and declaring AI use policies.

A distinct mode of reuse concerns our interpretation of the EU AI Act. In particular, our five-concept structure to determine Annex~III high-risk AI systems has been adopted by multiple studies, including a cluster of inter-connected papers by Nokia Bell Labs~\cite{bogucka2024co, herdel2024exploregen, rao2025riskrag, constantinides2024goodintentionsriskyinventions, bogucka2024ai, bogucka2024atlas, bogucka2025impact, vscepanovic2025ai, bogucka2024atlas}. A further paper from the lab  uses the definitions from AIRO and VAIR within prompts to guide LLMs to analyse harms associated with AI incidents \cite{velazquez2024decoding}.

%a prominate theme is Combination of knowledge graphs and LLMs

%Reuse of legal analysis specifically the 5 concepts for determing high-risk AI systems, As a guide for LLMs to do tasks -

Two industry-based research groups have reused our work: \textit{IBM research} reused AIRO in the risk Atlas Nexus~\cite{IBM_2026} and their subsequent work ~\cite{Daly2025usge_ibm, bagehorn2025airiskatlastaxonomy} and the \textit{Nokia Bell Labs} cluster of work mentioned above. This uptake outside academia is an indication of the usability of our artefacts in applied settings.

A further mode of reuse concerns our ontology development approach, specifically its iterative development and modular design, which is adopted across a series of publications by Russo~\cite{russo2024biaso, russo2025towards, russo2025documenting}. As these constitute a single evolving line of work, in our analysis we count them as one contribution.

%\textbf{Industry adoption} of particularly by IBM research in the risk Atlas Nexus \cite{IBM_2026} and the work build upon that ~\cite{Daly2025usge_ibm, bagehorn2025airiskatlastaxonomy} and a cluster of inter-connected papers by Nokia Bell Labs~\cite{bogucka2024co, herdel2024exploregen, rao2025riskrag, constantinides2024good, bogucka2024ai, bogucka2024atlas, bogucka2025impact, vscepanovic2025ai, bogucka2024atlas} reused the the 5 AIRO concepts structure we proposed for determining AI Act's high-risk AI system, which are \textit{domain}, \textit{purpose}, \textit{capability}, \textit{deployer}, and \textit{AI subject}.  Another paper from the same lab, 

Two studies assess our contributions:  \cite{Sooklall2025linguistic} evaluated VAIR directly and \cite{bogucka2024co} validated our five-concept structure in consultation with AI practitioners and AI compliance experts as part of assessing their proposed AI impact assessment report template.

%Advances the state of art by providing a set of open, common, interoperable, and extensible artefacts, upon which researchers can build specific risk ontologies and tools for conducting, documenting, and sharing AI risk and impact assessments. as shown in .....

%AIRO and VAIR can be leveraged for annotating AI use cases and incidents to classify, collate, and compare AI risks and risk assessment activities over time, as shown in \cite{}

%- Provided under open licenses and in a modular form, AIRO and VAIR can be enhanced and extended to meet the sector-specific requirements and provisions of the forthcoming guidelines and standards. Further, they can be expanded as per national AI policies and guidelines that might be issued by Member States. These extensions can co-exist under separate namespaces reusing the open, consistent, and standardised representation of information offered by AIRO and/or VAIR. In this case, a key advantage of utilising open knowledge modelling standards, as opposed to proprietary solutions, is in providing an open and transparent way to conduct mappings, assess alignment with the AI Act, and check their consistency. This is specially helpful in supporting the coordination among Member States and market surveillance authorities to tackle differences in interpretations of the AI Act—in particular those related to fundamental rights, that might disrupt the functioning of the EU digital single market.  

\section{Conclusion} \label{sec:conclusion}
This paper introduced AIRO and VAIR as ontologies for modelling AI use cases and their associated risk management information in a FAIR manner. AIRO and VAIR have been considered as the state of the art by providing novel AI risk ontologies and vocabularies based on the EU AI Act and ISO/IEC standards related to AI risks, including ISO/IEC 22989 \cite{iso22989}, ISO/IEC 23894 \cite{iso23894} and ISO 31073 \cite{iso31073}. 

In terms of limitations, the reliance of the ontologies on the interpretation of the AI Act and the relevant standards might limit their use. However, being implemented as open knowledge graphs, both ontologies can be adopted and enhanced freely through inheritance or concept specialisations. Further, we evaluated the ontologies using a limited set of illustrative examples, rather than comprehensive set of real-world use cases. While our evaluation was limited, the academic uptake of AIRO and VAIR and their reuse in industry settings provides complementary evidence of their real-world applicability. Additionally, the review of  AIRO and VAIR  by the experts within DPVCG during their integration into DPV, provided a further assessment of the ontologies by practitioners.

The overall approach towards the iterative design of AIRO and VAIR shows how ontologies can evolve in light of legal uncertainty. Moreover, the  modular approach in constructing consistent and integrable vocabularies for AI risks enables building the specific AI risk ontologies and taxonomies; such as sector- or technology-specific ones.

\section*{Funding}
This project has received funding from the European Union’s Horizon 2020 research and innovation programme under the Marie Skłodowska-Curie grant agreement No 813497 (PROTECT ITN), from the Horizon Europe Research and Innovation Programme under the FORSEE (grant agreement No 101177579) and RECITALS  (grant agreement No 101168490) projects, and from the ADAPT Centre for Digital Media Technology, funded by Research Ireland and is co-funded under the European Regional Development Fund (ERDF) through Grant\#13/RC/2106\_P2.

\section*{Declaration of generative AI and AI-assisted technologies in the manuscript preparation process}

During the preparation of this work the first author used Anthropic's Claude to assist with summarising citing publications in Section 9 and with language editing of the manuscript. All AI-assisted outputs were reviewed, verified against the primary sources, and corrected by the first author, who take full responsibility for the content of the publication.


\bibliographystyle{cas-model2-names}

\bibliography{bibliography}



\end{document}